\documentclass[article,notitlepage,preprintnumbers,superscriptaddress,nofootinbib,twocolumn]{revtex4-1}
\usepackage[utf8]{inputenc}
\usepackage[a4paper, total={7in, 9in}]{geometry}
\usepackage{amsmath}
\usepackage{amssymb}

\usepackage{enumerate}
\usepackage{tikz}
\usepackage[compat=1.1.0]{tikz-feynman}
\usepackage{feynmf}
\usepackage{slashed}
\usepackage{braket}
\usepackage{url}
\usepackage{natbib}
\usepackage{graphicx}
\usepackage[pdftex,bookmarks,linktocpage,pdfpagelabels,plainpages=false,hyperfigures,linkcolor=blue,citecolor=blue]{hyperref} 
\hypersetup{colorlinks=true}

\usepackage{mathrsfs}  
\usepackage{cancel}
\usepackage[normalem]{ulem}
\usepackage{caption}
\usepackage{subcaption}
\usepackage{multirow}

\begin{document}

\bibliographystyle{apsrev4-1}

\title{New Constraints on ALP Electron and Photon Couplings from ArgoNeuT and the MiniBooNE Beam Dump}

\author{Francesco Capozzi}  
\affiliation{Dipartimento di Scienze Fisiche e Chimiche, Università degli Studi dell’Aquila, 67100 L’Aquila, Italy}
\affiliation{Istituto Nazionale di Fisica Nucleare (INFN), Laboratori Nazionali del Gran Sasso, 67100 Assergi (AQ), Italy}

\author{Bhaskar Dutta}
\affiliation{Mitchell Institute for Fundamental Physics and Astronomy, Department of Physics and Astronomy, Texas A\&M University, College Station, TX 77845, USA}

\author{Gajendra Gurung}
\affiliation{Department of Physics, University of Texas, Arlington, TX 76019, USA}  

\author{Wooyoung Jang}
\affiliation{Department of Physics, University of Texas, Arlington, TX 76019, USA}  

\author{Ian M. Shoemaker}
\affiliation{Center for Neutrino Physics, Department of Physics, Virginia Tech, Blacksburg, VA 24061, USA}

\author{Adrian Thompson}
\affiliation{Mitchell Institute for Fundamental Physics and Astronomy, Department of Physics and Astronomy, Texas A\&M University, College Station, TX 77845, USA}

\author{Jaehoon Yu}
\affiliation{Department of Physics, University of Texas, Arlington, TX 76019, USA}  

\begin{abstract}
Beam dumps and fixed-target experiments have been very sensitive probes of such particles and other physics beyond the Standard Model (BSM) by considering the production of new states from the primary interaction in the beam dump. In a proton beam dump, there are many secondary interactions taking place in electromagnetic showers which may be additional production channels for pseudoscalar bosons or axion-like particles (ALPs). The target-less configuration of the MiniBooNE experiment, which collected data from $1.86 \times 10^{20}$ protons impinging directly on the steel beam dump, is an excellent test of sensitivity to these production channels of ALPs in the MeV mass region. Using the null observation of the MiniBooNE dump mode data, we set new constraints on ALPs coupling to electrons and photons produced through a multitude of channels and detected via both scattering and decays in the MiniBooNE detector volume. We find that the null result rules out parameter space that was previously unconstrained by laboratory probes in the 10-100 MeV mass regime for both electron and photon couplings. Lastly, we make the case for performing a dedicated analysis with 1.25$\times 10^{20}$ POT of data collected by the ArgoNeuT experiment, which we show to have complementary sensitivity and set the stage for future searches.
\end{abstract}

\preprint{MI-HET-808}

\maketitle

\section{Introduction}

Particle beam dumps have proven to be ultra-sensitive probes of new physics sectors beyond the Standard Model (BSM), where the myriad electromagnetic and hadronic cascades produce showers of electrons, positrons, gamma rays, and mesons; each a potential channel for BSM particle production. Studying the beam target environment and the particle showers within is thus a crucial first step to understanding what kind of physics is possible, and at what energy scales. Already many searches have been performed by electron beam dumps (E137, NA64, E141, Orsay, E774, etc.~\cite{Bechis:1979kp,PhysRevD.38.3375,Andreas:2010ms,Riordan:1987aw,Bross:1989mp,NA64_missingenergy,Andreev:2021fzd,Gninenko:2017yus}) and proton beam dumps at the GeV energy scale (e.g. CHARM, NuCal, NA62, SeaQuest/SpinQuest~\cite{Dobrich:2018ezn,Berlin:2018pwi,Tsai:2019buq,Andreas:2010ms}) and sub-GeV sources (e.g. CCM~\cite{CCM:2021lhc}, IsoDAR~\cite{Alonso:2021kyu}, and COHERENT~\cite{COHERENT:2021pvd}), and others~\cite{Agrawal:2021dbo}. 

The existence of pseudoscalar bosons with small couplings to the SM are predicted in models of broken symmetries in connection with explaining many puzzles in nature. Axions and axion-like particles (ALPs) are central features in the landscape of solutions, in particular, to the strong CP problem~\cite{Peccei:1977hh,Wilczek:1977pj,Weinberg:1977ma,Preskill:1982cy,Abbott:1982af,Dine:1982ah,Duffy:2009ig,Marsh:2015xka,Battaglieri:2017aum} and to the dark matter problem~\cite{Panci:2022wlc, Marsh:2015xka, Chadha-Day:2021szb}, and otherwise appear ubiquitously in string theory~\cite{Svrcek:2006yi, Halverson:2019cmy}, and the ultraviolet spectra of many other puzzle-solving models with spontaneously broken symmetries. In many of these scenarios, it is possible that the ALP has couplings to SM leptons and the electromagnetic field, making the particle showers inside the beam target good laboratory probes of ALPs, reaching up to GeV mass scales. ALPs at the MeV to GeV mass scales are of particular interest to beam dump and fixed target experiments and have been studied in the context of heavy axions~\cite{Kelly:2020dda,Gaillard:2018xgk, Kivel:2022emq, Hook:2019qoh}, whose parameter space extends beyond that of traditional QCD axion models.

In 2018 MiniBooNE collaboration performed an analysis of their targetless-mode run~\cite{Aguilar-Arevalo:2018wea}, in which they collected data associated with $1.86 \times 10^{20}$ protons on target (POT) bypassing the main beryllium target and impinging on the steel beam dump. Expected neutrino rates for this mode were very low, and no excess of events was observed, in contrast to the results from the target-mode runs~\cite{MiniBooNE:2018esg,MiniBooNE:2020pnu}. In this work, we show that the null result from this data set is sensitive enough to ALPs produced in electromagnetic showers in the dump to set new limits on photon and electron couplings. 

Running in a target-less mode has the effect of suppressing the fluxes of neutrinos coming from charged meson decays. Searches for BSM particles that have production channels orthogonal to the charged pion decay gain a big advantage here; in the case of a thin target, the charged mesons decay in flight after getting produced, allowing them to be focused by the magnetic horn system. In the thick beam dump case, however, the charged pions are stopped in the material and decay isotropically, suppressing the subsequent neutrino background that would lie in the signal region for the BSM search.

This realization is especially important for future beam dump experiments at higher energies, where the higher intensity of electromagnetic cascades provide both the coupling and mass reach necessary to significantly extend the limits tested so far by laboratory searches in the MeV to GeV mass range. We will show that data collected by the ArgoNeuT detector~\cite{ArgoNeuT:2018tvi,ArgoNeuT:2022mrm} already has this capability, and depending on the specific sensitivity of a dedicated analysis, null observations in this data could already rule out parameter space unconstrained by laboratory probes to-date.

In \S~\ref{sec:production} we outline the production and detection channels we consider for electromagnetically-coupled ALPs. In \S~\ref{sec:analysis} we describe the statistical analysis performed for the MiniBooNE dump-mode data and the ArgoNeuT data given an ALP signal hypothesis, with the resulting limits placed on the parameter space of photon and electron couplings in \S~\ref{sec:results}. Finally we conclude in \S~\ref{sec:outlook}.

\section{BSM Production and Detection in a Beam Dump}
\label{sec:production}

We consider primarily ALPs produced in electromagnetic cascades inside the beam dump or beam target environment, e.g., those that get produced from couplings to photons and to electrons;
\begin{equation}
    \mathscr{L}_{ALP} \supset i g_{ae} a \bar{\psi}_e \gamma^5 \psi_e - \frac{1}{4} g_{a\gamma} a F_{\mu\nu}\Tilde{F}^{\mu\nu}
\end{equation}
This Lagrangian, which for simplicity we will assume only one tree-level coupling active or dominant at a time, opens up a slew of production and detection channels available to beam target and beam dump experiments. These have recently been investigated in refs.~\cite{Dent:2019ueq, Brdar:2020dpr, Dutta:2020vop, Capozzi:2021nmp,CCM:2021lhc, Brdar:2022vum}, and we summarize them in Table~\ref{tab:channels}.
\begin{table}[b!]
    \centering
    \begin{tabular}{|c|c|c|}
        \hline
         Coupling & Production & Detection \\
         \hline
         \multirow{2}*{$g_{a\gamma}$} & \multirow{2}*{$\gamma \, Z \to a \, Z$} & $a \, Z \to \gamma \, Z$ \\
         & & $a \to \gamma \gamma$ \\
         \hline
         \multirow{5}*{$g_{ae}$} & $\gamma \, e^- \to a \, e^-$ & \multirow{5}*{$a \, e^- \to \gamma \, e^-$} \\
         & $e^+ \, e^- \to a \, \gamma$ & \multirow{5}*{$a \to e^+ e^-$} \\
         & $e^\pm \, Z \to e^\pm \, Z \, a$ & \multirow{5}*{$a \, Z \to e^+ \, e^- \, Z$}  \\
         & $e^+ \, e^- \to a$ & \\
         & $\pi^\pm \to e^\pm \, \nu \, a$ & \\
         \hline
    \end{tabular}
    \caption{ALP production and detection mechanisms that are available through couplings to electrons and photons considered in this analysis.}
    \label{tab:channels}
\end{table}

For ALPs coupled to electrons, the dominant final state will be $e^+ e^-$ pairs appearing in the detector as single Cherenkov rings, either from the pair being highly collinear with a separating angle less than the typical angular resolution of the detector or if one of the electrons/positrons are too soft. This final state appears mainly through decays for $m_a > 2 m_e$ and otherwise through the Bethe-Heitler lepton pair production process ($a Z \to e^+ e^- Z$) for sub-MeV ALPs, considered before to set limits on light (pseudo)scalars appearing in a proton beam target~\cite{Blumlein:1990ay,Blumlein:1991xh}. The cross-section for this process was computed in refs.~\cite{KIM198387, KIM1984189} using the formalism and atomic form factors presented in ref.~\cite{RevModPhys.46.815}, and it is larger than inverse-Compton scattering ($a e^- \to \gamma e^-$) by up to an order of magnitude for ALP energies in the 100 MeV - 1 GeV range, which is the energy region of interest for this study.

The resonant cross section in the electron rest frame is
\begin{align}
    \sigma &= \dfrac{2 \pi m_e g_{ae}^2 s}{m_a^2 \sqrt{s(s-4m_e^2)}} \delta(E_+ - (\frac{m_a^2}{2m_e} - m_e)) \\
    &\simeq \dfrac{2\pi m_e g_{ae}^2}{m_a^2}\delta(E_+ - (\frac{m_a^2}{2m_e} - m_e)).
\end{align}

\begin{figure}[h]
    \centering
    \includegraphics[width=\linewidth]{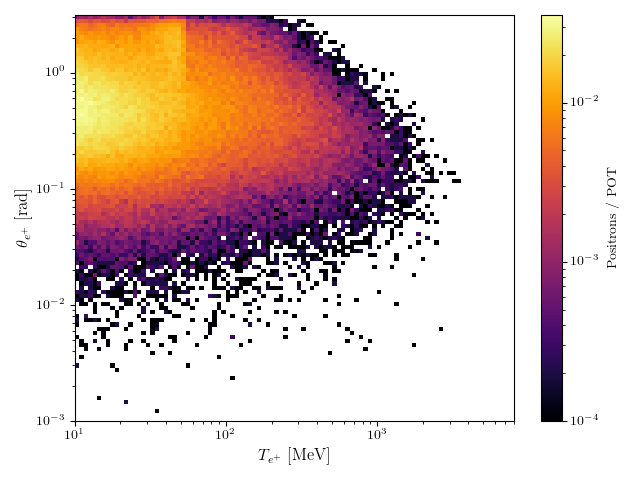}
    \includegraphics[width=\linewidth]{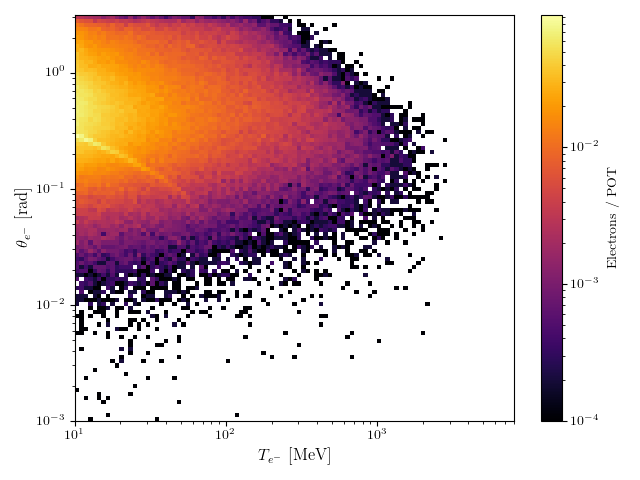}
    \caption{Electron and positron fluxes in the BNB steel beam dump generated with GEANT4 using $10^4$ POT with the \texttt{QGSP\_BERT\_AllHP} physics list, here shown as a function of the $e^\pm$ angle with respect to the beam axis and kinetic energy $T$.}
    \label{fig:pos_flux_2d}
\end{figure}

\begin{figure}[h]
    \centering
    \includegraphics[width=\linewidth]{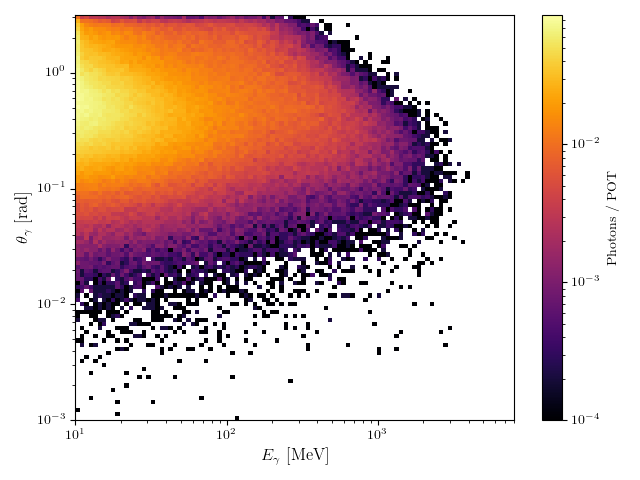}
    \caption{Photon fluxes in the BNB steel beam dump generated with GEANT4 using $10^4$ POT with the \texttt{QGSP\_BERT\_AllHP} physics list, here shown as a function of the $\gamma$ angle with respect to the beam axis and energy $E_\gamma$.}
    \label{fig:photon_flux_2d}
\end{figure}

\begin{figure}[h]
    \centering
    \includegraphics[width=\linewidth]{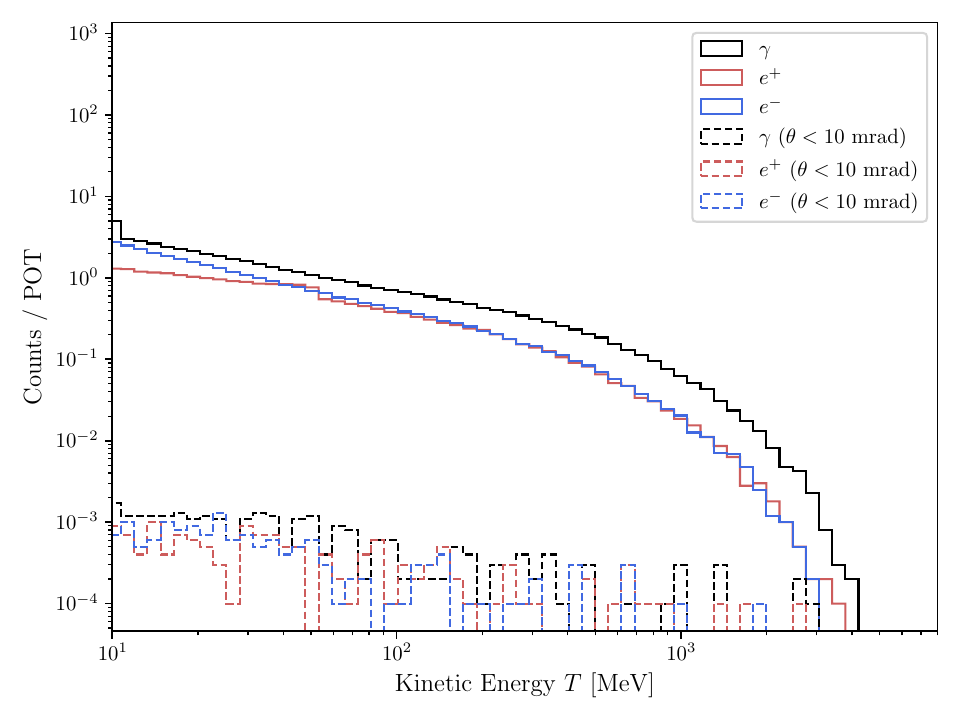}
    \caption{Electron, positron, and photon fluxes before and after angular cuts in the MiniBooNE beam dump.}
    \label{fig:epem_flux}
\end{figure}

To simulate the production fluxes, we first generate the SM particle fluxes inside the MiniBooNE dump with GEANT4 using the \texttt{QGSP\_BERT\_AllHP} physics list, then pass a high-statistics sample of each particle flux ($e^\pm, \gamma, \pi^\pm$) into the \texttt{alplib} event generator.\footnote{\href{https://github.com/athompson-git/alplib}{https://github.com/athompson-git/alplib}}
The positron and electron fluxes are shown in Fig.~\ref{fig:pos_flux_2d}, while the photon flux is shown in Fig.~\ref{fig:photon_flux_2d}. We show a large phase space of the $e^\pm$ and $\gamma$ fluxes to illustrate the many low-energy features that come about from processes like nuclear de-excitation and beta decay. However, in principle, only the high energy tail ($>75$ MeV) in the forward-going region ($\theta \lesssim 10^{-2}$ rad) is responsible for the bulk of BSM particle production that is captured within the signal region and pointing within the solid angle of the MiniBooNE detector. This is illustrated in Fig.~\ref{fig:epem_flux} where we show the energy spectra before and after an angular cut of 10 mrad. Further details of the event selection and signal window are discussed in the following section.

For ALPs produced from electrons or positrons in resonant production ($e^+ e^- \to a$), associated production ($e^+ e^- \to a \gamma$), or bremsstrahlung ($e^\pm Z \to e^\pm Z a$), the energy loss of the electrons and positrons in the material during particle transport must also be folded into the event rate calculation. This modifies the number flux leaving the beam dump as 
\begin{align}
\label{eq:alp_flux}
    \dfrac{dN_a}{dE_a} &= \frac{N_A X_0}{A} (\hbar c)^2 \int \frac{d^2\Phi_{e^+}}{dE_e d\Omega_e}  I(t, E_+, E^\prime) \nonumber \\ &\times  \Theta_\text{det} \dfrac{d^2\sigma(E^\prime)}{dE^\prime d\Omega^\prime} d\Omega_e d\Omega^\prime dE_+ dt dE^\prime 
\end{align}
where $N_A$ is Avogadro's number, $X_0$ is the radiation length of the electrons/positrons in the dump material, and $A$ is the atomic weight. $I(t, E_i, E_f) = \frac{\theta(E_i - E_f)}{E_i \Gamma (4 t/3)} (\ln E_i/E_f)^{4t/3 - 1}$ is the energy loss smearing function for the electron/positron radiation length $t$ integrated up to target radiation thickness $T$~\cite{PhysRevD.34.1326}. We integrate over the solid angle of the positron with respect to the beamline, $\Omega_e$, and outgoing ALP solid angle with respect to the positron direction, $\Omega^\prime$, taking care to integrate only those ALPs pointed in the direction of the detector solid angle through the Heaviside function $\Theta_\text{det}$~\cite{Brdar:2020dpr}.


\section{Data Analysis}
\label{sec:analysis} 
\subsection{MiniBooNE Dump Mode}
\begin{figure}[!t]
    \centering
    \begin{subfigure}{.52\textwidth}
        \includegraphics[width=\textwidth]{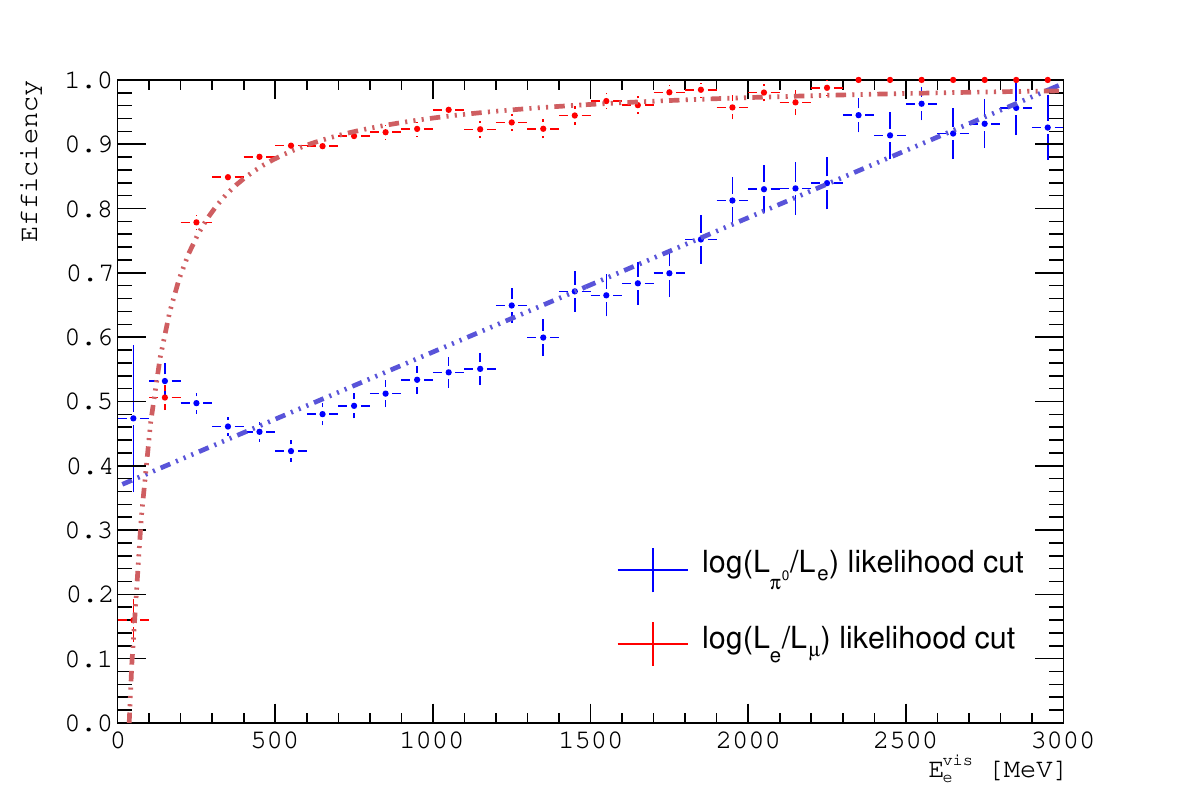}
        \caption{Selection efficiency as a function of $E^{vis}_{e}$.}
        \label{Fig: EnergyDist}
    \end{subfigure}
    \begin{subfigure}{.52\textwidth}
    \centering
        \includegraphics[width=\textwidth]{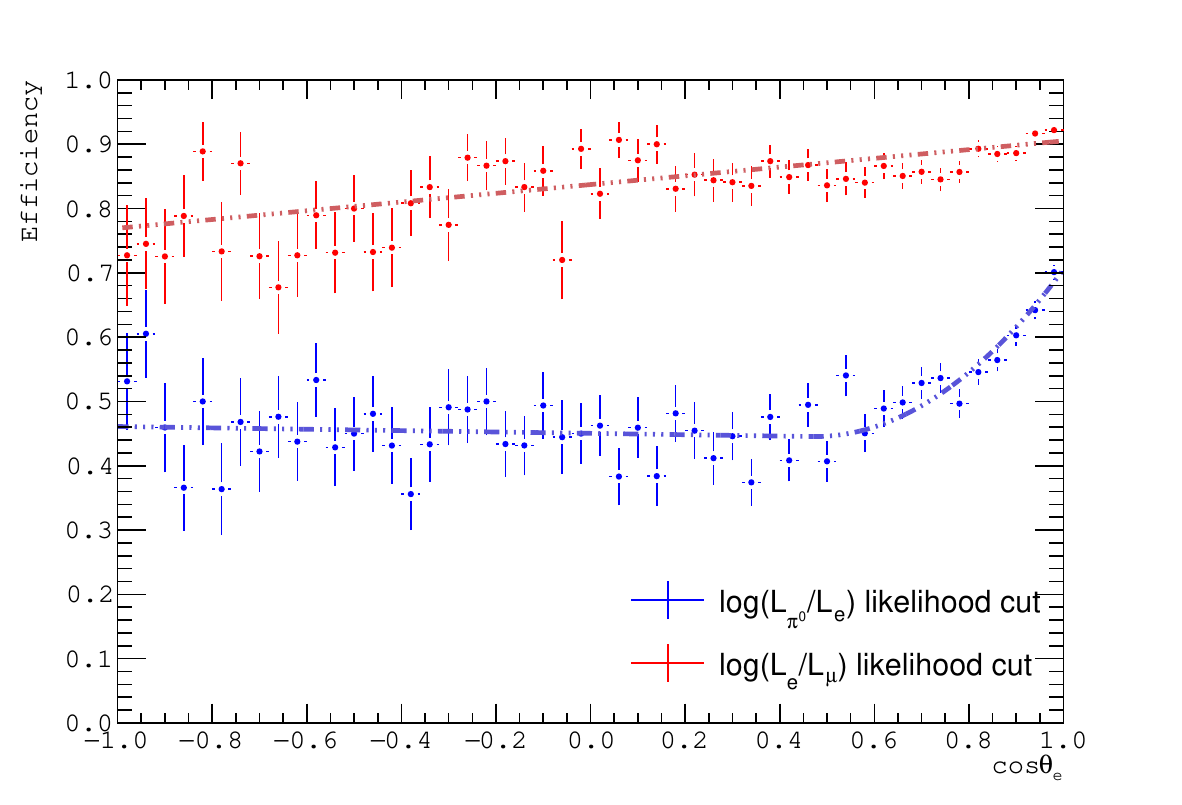}
        \caption{Selection efficiency as a function of $\cos\theta_{e}$.}
        \label{Fig: AngDist}
    \end{subfigure}
    \caption{Selection efficiency of the electron-like events from reconstructed events inside the MiniBooNE detector.}
    \label{fig:efficiencies}
\end{figure}

The final states of concern in our search for ALPs in the MiniBooNE detector are photon-like events and electron-like events, listed in Table~\ref{tab:channels}. We have adopted the same selection cuts made in the $\nu-e$ analysis of the MiniBooNE dump mode data for these states. Here we study the detector response with true simulated information to analyze the efficiency of the electron-like event selection from reconstructed events inside the detector. For the analysis of the Monte Carlo generated data, after the preliminary cuts have been applied, the first round of the reconstructed events is fit under the one-track electron and muon hypothesis. Each fit returns the likelihood of the corresponding hypothesis: $\mathcal{L}_e$ and $\mathcal{L}_\mu$. Those events satisfying the $\log(\mathcal{L}_e/\mathcal{L}_\mu) > -0.05$ continue the next round of reconstruction. In the second round, reconstructed events are fit under the general two-photon hypothesis. Similarly, the events should satisfy $\log(\mathcal{L}_{\pi^0}/\mathcal{L}_e) < 0$. The efficiencies of these two cuts using simulated data as functions of electron visible energy and electron scattering angle are shown in Fig.~\ref{fig:efficiencies}. The selection efficiencies as a function of the visible energy, $E^{vis}_{e}$, are fitted as an arctangent function ($p_0\arctan(p_1 x) + p_2$). The selection efficiencies as a function of the cosine of the angle with respect to the beam axis, $\cos\theta_{e}$, are fitted as a straight line ($p_0+p_1x$) except for the forward region of $\log(\mathcal{L}_e/\mathcal{L}_\mu)$ which has a second-order polynomial fit ($p_0+p_1x+p_2x^2$). Uncertainties from the goodness-of-fit on the efficiency curve as a function of $E_e^{vis}$ and $\cos\theta_e$ are constrained to be less than $20\%$, so their impact on the exclusions over the model parameter space shown in the following section will not be qualitatively different. 

In addition to these log-likelihood efficiencies, we also take into account the cut on the reconstructed vertex radius of 500 cm, which effectively reduces the MiniBooNE volume to a sphere of 10 m in diameter. Other cuts, such as the number of tank and veto hits, and the Scintillation / Cherenkov ratios we assume to have perfect signal efficiency for the detection channels in Table~\ref{tab:channels}. However, we do check that the $\gamma \gamma$, $e^+ e^-$, and $\gamma e^-$ final states from axion interactions and decays are collinear enough to be identified as a single electron-like Cherenkov ring in the detector. This also ensures that the cut on the di-gamma invariant mass $m_{\gamma\gamma} \leq 80$ MeV is passed by selection for our ALP signals.

\begin{figure*}[!ht]
    \centering
    \includegraphics[width=0.24\textwidth]{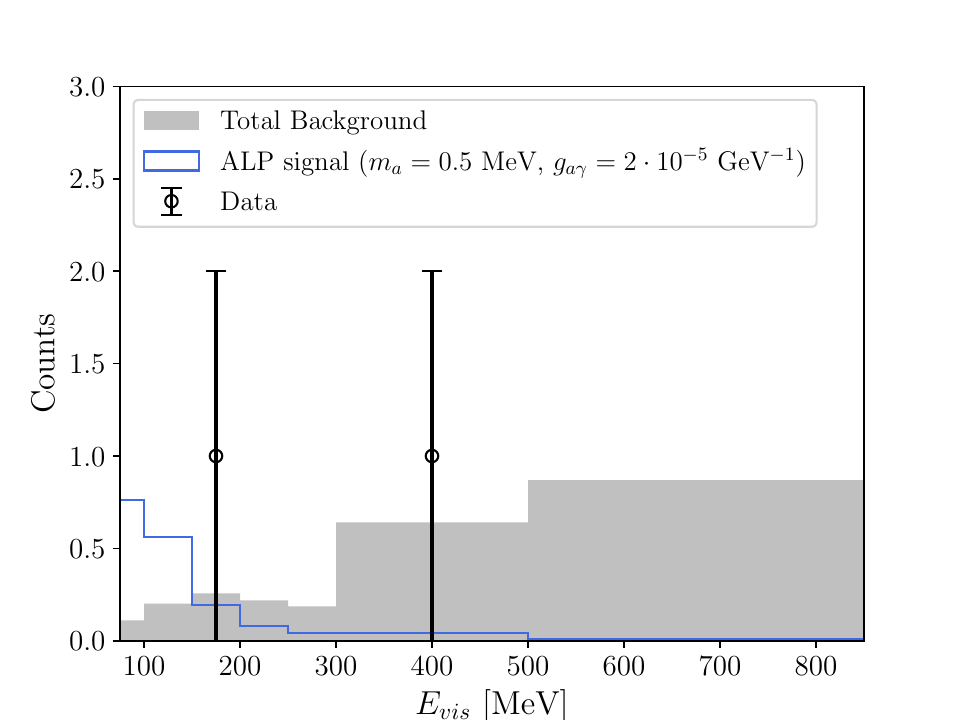}
    \includegraphics[width=0.24\textwidth]{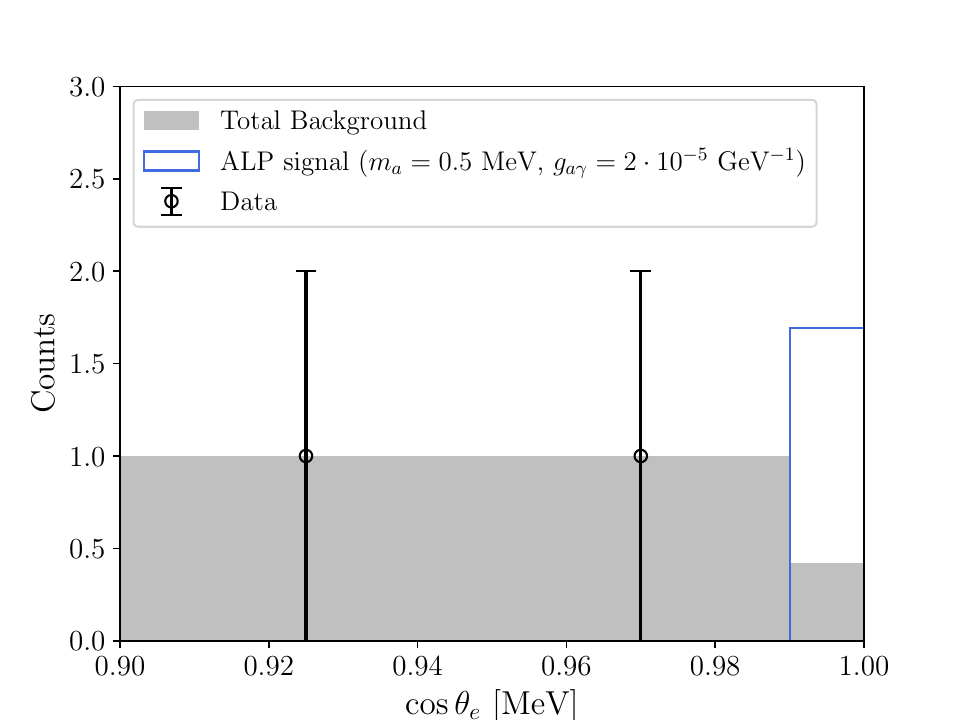}
    \includegraphics[width=0.24\textwidth]{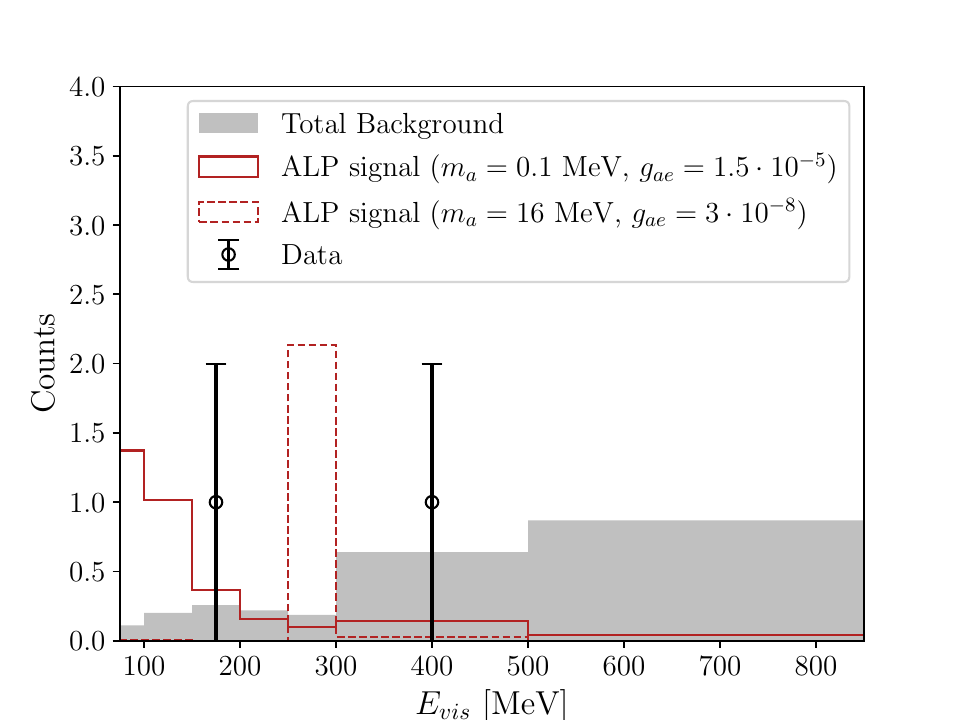}
    \includegraphics[width=0.24\textwidth]{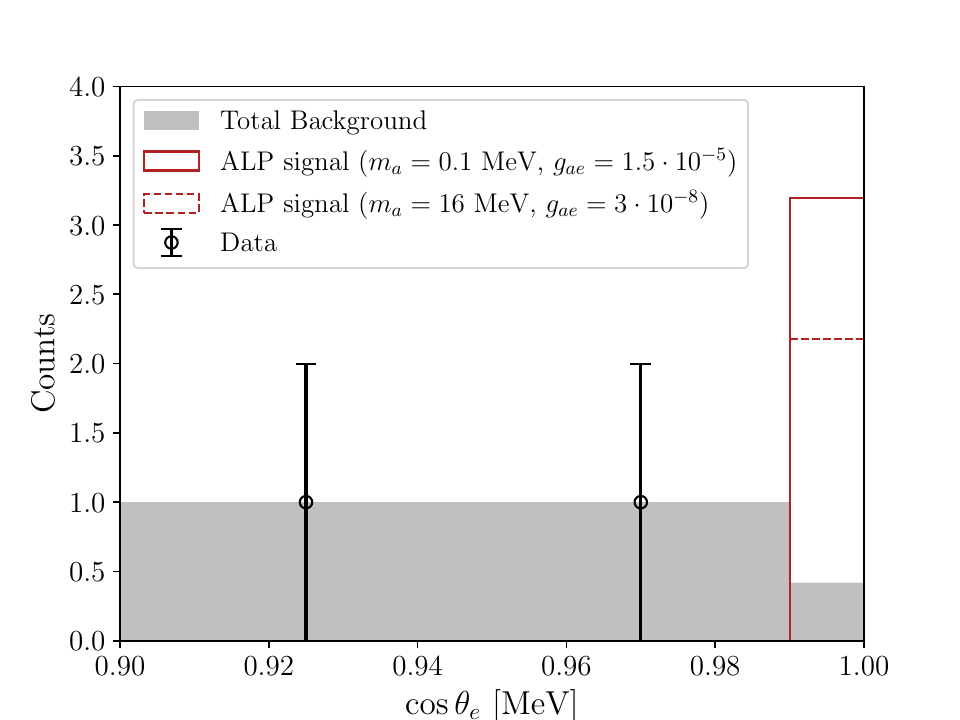}
    \caption{Left: visible energy spectrum from ALP production and scattering in photon coupling channels, where a mass of $m_a=0.5$ MeV at couplings of $g_{a\gamma} \sim \mathcal{O}(10^{-5})$ GeV$^{-1}$ mainly contribute to the decay channel $a\to\gamma\gamma$.
    Right: visible energy spectra from electron coupling channels, with the solid red histogram showing the predicted spectra below the decay threshold $m_a = 2 m_e$, while the dashed red histogram shows the spectrum for a larger mass which is resonantly produced and dominantly produces a $a \to e^+ e^-$ decay signature.}
    \label{fig:evis_spectra}
\end{figure*}

Lastly, we bin the ALP signal Monte Carlo events into visible energy and cosine bins between $75 \leq E_\gamma \leq 850$ MeV and $\cos\theta \geq 0.9$ (taking $E_\gamma = E_e^{vis}$ for the electron-like visible energy measurement). Since inverse Primakoff scattering is characterized by a forward outgoing photon, while inverse Compton scattering is characterized by a forward outgoing electron and a soft off-forward photon (typically below the lower energy cut), these scattering channels are well within the selection region for most choices of the couplings and the ALP mass. Example spectra for photon and electron coupling channels are shown in Fig.~\ref{fig:evis_spectra}, where we have convolved the predicted event rates with the efficiency functions described above. For the case of ALPs undergoing inverse Primakoff scattering in the detector, $a Z \to \gamma Z$, we integrate over the visible energy and outgoing angle of the final state photon;
\begin{equation}
\label{eq:evt_rate}
    \frac{d^2R}{dE_\gamma d\Omega_\gamma} = N_T \int \frac{dN_a}{dE_a} \frac{d^2\sigma(E_a)}{dE_\gamma d\Omega_\gamma} \epsilon(E_\gamma) \epsilon(\Omega_\gamma) dE_a
\end{equation}
where $\epsilon(E_\gamma)$ and $\epsilon(\Omega_\gamma) = \epsilon(\cos\theta_\gamma)$ are equivalent to the visible energy and cosine efficiencies, respectively, of the electron-like signals shown in Fig.~\ref{fig:efficiencies}. Here, recall the differential event rate $dN_a / dE_a$ passing into the detector from Eq.~\ref{eq:alp_flux}.

Integrating Eq.~\ref{eq:evt_rate} over energy bin edges $[75, 100, 150, 200, 250, 300, 500, 850]$~(in MeV) and cosine bin edges $[0.9, 0.95, 0.99, 1.0]$ yields the ALP signal $s_i$ in each bin $i$ as a function of the mass and couplings. In the case of decays, instead of the differential cross section in Eq.~\ref{eq:evt_rate} we use the probability of decays occurring inside the detector
\begin{equation}
    P_{decay}= e^{-\ell/(\tau v_a)} \left[ 1 - e^{-\Delta\ell /(\tau v_a) } \right]\,
\end{equation}
where $\tau v_a$ is the ALP decay length in the lab frame, $\ell$ is the baseline distance between the ALP production in the dump, and $\Delta \ell$ is the fiducial path length in the detector during which the decay must take place.

For the other detection channel final states ($2\gamma$, $1\gamma1e^-$, or $e^+e^-$), both final state particles leave visible energy in the detector, so we need to ensure that they are collinear enough to be reconstructed as a single Cherenkov ring in the detector. We check the angular distribution of the final state and cut events if two final state particles are separated by more than 5 degrees.

We use a binned log-Poisson likelihood to obtain the confidence limits;
\begin{equation}
   \ln L(\vec{\theta}) = \sum_{i=1}^7 d_i \ln \big[s_i(\vec{\theta}) + b_i\big] - \big[s_i(\vec{\theta}) + b_i\big] - \ln \big[\Gamma(d_i + 1)\big]    
\end{equation}
for data $d_i$, backgrounds $b_i$, and signal $s_i(\vec{\theta})$, where $\vec{\theta} = (m_a, g_{a\gamma})$ in the case of dominant ALP-photon coupling and $\vec{\theta} = (m_a, g_{ae})$. The CLs are then given by finding regions of constant delta-log-likelihood,
\begin{equation}
    -2\Delta \ln L \equiv 2(\ln L(\theta) - \ln L(\theta)_{min}),
\end{equation}
in the relevant model parameter space $\vec{\theta}$.

\subsection{ArgoNeuT}
\begin{figure}
    \centering
    \includegraphics[width=0.5\textwidth]{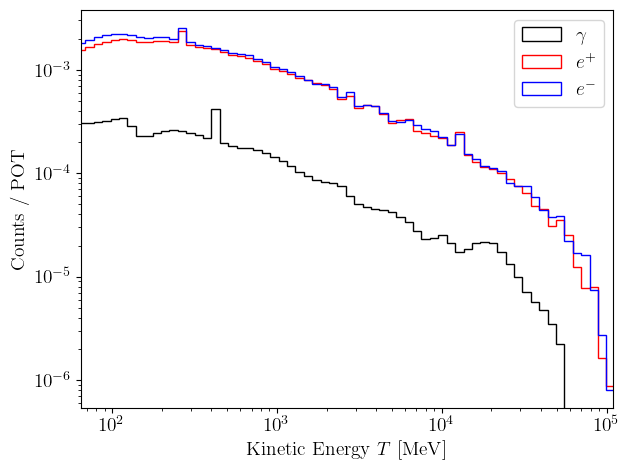}
    \caption{Electron, positron, and photon fluxes at the NuMI target whose directions point within the solid angle of the ArgoNeuT detector.}
    \label{fig:argoneut_fluxes}
\end{figure}

ArgoNeuT~\cite{ArgoNeuT:2018tvi} collected data from $1.25 \times 10^{20}$ POT impinging on the NuMI target, with its LArTPC detector situated 1.04 km downstream of the target while the beamline was in anti-neutrino mode~\cite{Adamson:2015dkw}. With a fiducial volume of $0.40\times0.47\times0.90$ cm$^3$, the angular acceptance of the detector coverage corresponds to roughly 0.325 mrad in solid angle. We perform a similar simulation with GEANT4 using the \texttt{QGSP\_BERT\_AllHP} physics list to model the particle cascades inside the NuMI beam target environment (120 GeV protons on graphite). The ALP flux is calculated in the same way explained in the case of the MiniBooNE dump. From the GEANT4 flux distributions of $e^\pm$ and $\gamma$ in the solid angle of ArgoNeuT, shown in Fig.~\ref{fig:argoneut_fluxes}, we estimate the ALP flux produced from 1.25$\times 10^{20}$ POT during data collection.

A dedicated search for heavy ALPs decaying to di-muon pairs was performed by the ArgoNeuT collaboration~\cite{ArgoNeuT:2022mrm}, exhibiting an event topology with very low background expectations. However, here we are interested in different types of event topologies: $e^+ e^-$, $e^- \gamma$, $2\gamma$ and $1\gamma$ (see Table~\ref{tab:channels}), for which a dedicated analysis is missing. Therefore, we will not perform a likelihood analysis. We will just provide the contours in the parameter space for which the following number of signal ALP events would be observed in ArgoNeuT: 3, 20, and 100. These numbers are equal to the Poisson error of $\sim $ 10, 400, and $10^4$ background events, respectively.

\section{Results}
\label{sec:results}
\begin{figure}
    \centering
    \includegraphics[width=0.5\textwidth]{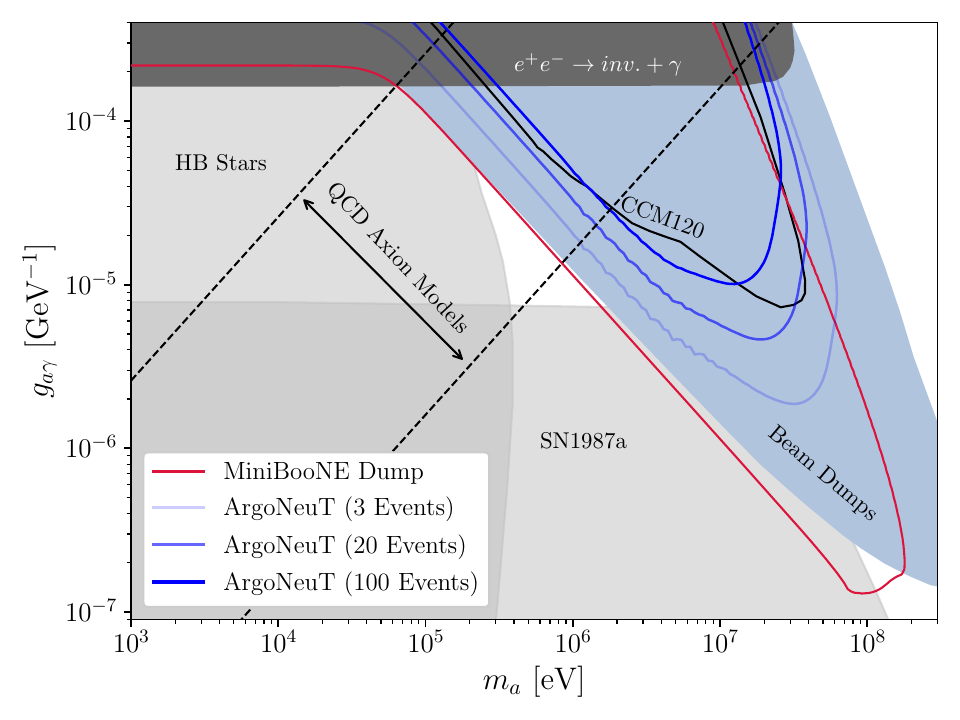}
    \caption{Constraints on $g_{a\gamma}$ set by the MiniBooNE dump-mode run at $2\sigma$ C.L. are shown, while the expected limits set by ArgoNeuT are shown for several benchmark signal rates.}
    \label{fig:photon_limits}
\end{figure}

The constraints on the ALP-photon coupling $g_{a\gamma}$ as a function of the ALP mass $m_a$ derived from MiniBooNE beam dump mode data is shown in Fig.~\ref{fig:photon_limits}. The $1\sigma$ and $2\sigma$ CLs are shown individually using the delta-log-likelihood method, and we find that the MiniBooNE data sets new laboratory limits on the ALP coupling for masses below 100 keV or so, where previously astrophysics (HB star cooling and SN1987a~\cite{Lucente:2020whw}, see also refs.~\cite{Jaeckel:2006xm,Khoury:2003aq,Masso:2005ym,Masso:2006gc,Dupays:2006dp, Mohapatra:2006pv,Brax:2007ak,DeRocco:2020xdt}) had placed the only constraints ahead of beam dump constraints~\cite{Jaeckel:2015jla,doi:10.1142/S0217751X9200171X}\footnote{The measurement of the explosion energy
of SN1987A can have  tension to the cosmological triangle region unless the star cooling process is significantly different
from the standard picture~\cite{Caputo:2021rux}.} and recently, constraints set by the CCM120 engineering run~\cite{CCM:2021lhc}. Limits set by the ArgoNeuT null result from 1.25$\times 10^{20}$ POT of collected data are shown in blue, benchmarking the signal event rate at 3, 20, and 100 events in the absence of a dedicated analysis with backgrounds and proper event selection. Comparing the shape of the exclusion contours between MiniBooNE and ArgoNeuT, one can see the impact of the longer baseline between beam target and detector at ArgoNeuT ($\sim 1$ km) versus MiniBooNE ($489$ m) shifting the sensitivity contour to larger masses reflecting longer ALP lifetimes for $a \to \gamma \gamma$ decay.

In this space, we also show the parameter space associated with QCD Axion model benchmarks spanned between the dashed black lines. Here the range of couplings and masses are shown for Kim-Shifman-Vainshtein-Zakharov (KSVZ) benchmark models~\cite{PhysRevLett.43.103, SHIFMAN1980493}, where the range is defined by taking the anomaly number ratios of $E/N = 44/3$ to $E/N = 2$ in the model. The correlations between the QCD axion mass and its effective couplings are taken from ref.~\cite{DiLuzio:2020wdo} (see also Appendix~\ref{app:qcd}). While the constraints shown here are purely on the photon-ALP couplings, independent constraints on the ALP-gluon couplings in these model variants are stringent and would indirectly rule out much of the parameter space~\cite{Alonso-Alvarez:2018irt}. These bands are of course only representative of these traditional QCD models shown for a sense of scale. QCD axions that are invoked to solve the strong CP problem which have parametrically heavier or lighter masses in other non-traditional models are also possible~\cite{Elahi:2023vhu, Gaillard:2018xgk, Kivel:2022emq, Hook:2019qoh}.

\begin{figure}
    \centering
    \includegraphics[width=0.5\textwidth]{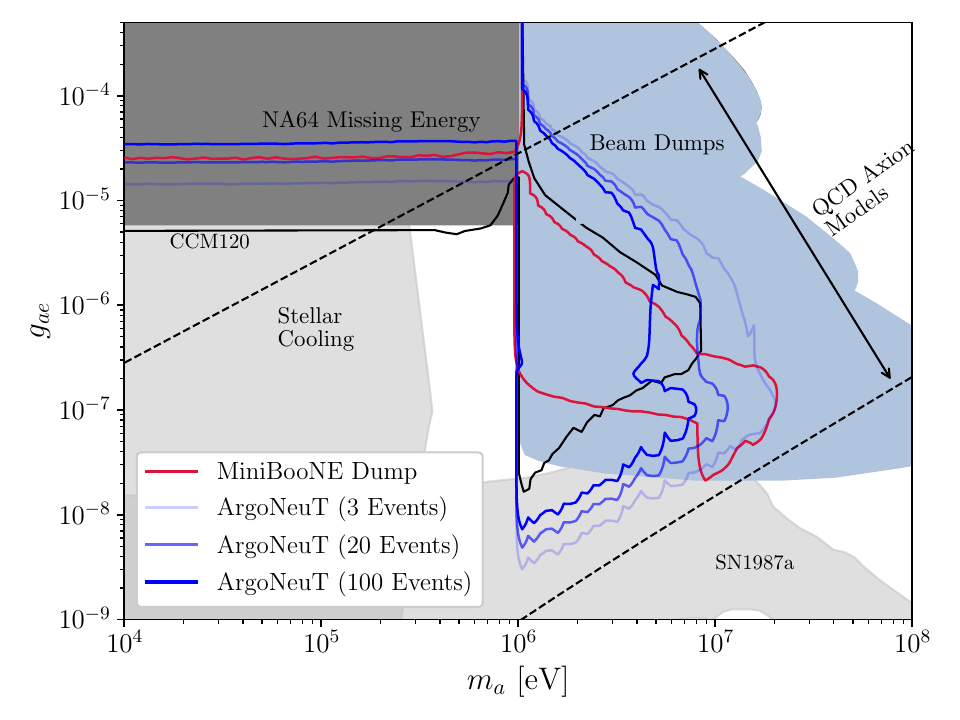}
    \caption{Constraints on $g_{ae}$ set by the MiniBooNE dump-mode run at $2\sigma$ C.L. are shown, while the expected limits set by ArgoNeuT are shown for several benchmark signal rates.}
    \label{fig:electron_limits}
\end{figure}

We set limits in the same way on the electron-ALP coupling $g_{ae}$ as a function of the ALP mass in Fig.~\ref{fig:electron_limits}. The parameter space associated with Dine-Fischler-Srednicki-Zhitnitsky (DFSZ) benchmark models~\cite{Zhitnitsky:1980tq,DINE1981199,Dine:1981rt,Dine:1982ah}, for which couplings to electrons would be dominant relative to the photon couplings, the span between the dashed black lines. Again, we show this span of model parameter space for reference although the constraints shown here from pure $g_{ae}$-driven channels are conservative and indirect constraints on the DFSZ gluon couplings would be more stringent. In the electron coupling, we find that MiniBooNE dump mode tests parameter space already ruled out by existing laboratory searches (e.g. NA64, E137, and other beam dumps). Although, in the mass range $\sim 10$ MeV the resonant channel $e^+ e^- \to a$ produces a highly peaked signal which becomes visible inside the energy region of interest, $75 < E_{vis} < 850$ MeV (see Fig.~\ref{fig:evis_spectra}). This is because the resonant energy tracks the square of the ALP mass, as $E_a = m_a^2 / (2m_e)$, producing the first visible peak within this energy range for $m_a \simeq 10$ MeV. The MiniBooNE dump mode becomes highly sensitive to ALP signals here for those masses but is consistent with the existing E137 constraints in this region. The subtle undulating features in the CL contours from $m_a = 10 - 30$ MeV then reflect the signal rising and falling to accommodate the two data points in the 3rd and 6th energy bins in Fig.~\ref{fig:evis_spectra}.

ArgoNeuT sensitivity to this coupling is fairly powerful in the $m_a > 2m_e$ mass range and would exclude new parameter space ahead of the limits set by the CCM120 engineering run between $m_a = 1$ MeV and $m_a = 5$ MeV. This is owed in part to the energy scale and long distance from the detector to the target being ideal to probe long ALP lifetimes, and also the relatively larger $e^\pm$ fluxes produced in the NuMI target (Fig.~\ref{fig:argoneut_fluxes}). This exclusion would be possible even for a benchmark signal rate of 100 events, corresponding roughly to a Poisson background of $10^4$ events without taking into account signal efficiency. This sensitivity is lost in the scattering limit for $m_a < 2 m_e$ where NA64 missing energy and CCM120, where being at much closer proximity to the production site, $\ell \sim 20$ m plays a bigger role, set the leading constraints.

\section{Outlook}
\label{sec:outlook}
The analysis of the MiniBooNE dump mode data shows significant sensitivity to dark sector states produced by the secondary electromagnetic cascades in the BNB dump environment. By utilizing the off-target configuration and examining the interactions of $1.86 \times 10^{20}$ protons with the steel beam dump, we have expanded the existing constraints on ALPs in the 10-100 MeV mass regime that couple to photons. Simultaneously, despite a small exposure and fiducial detector mass, the null observations of ArgoNeuT could potentially rule out parameter space for ALPs in the same mass range coupling to electrons, due to the higher beam energy.

Stopped-pion experiments at $\sim$GeV scale proton beam dumps also have the capability to probe new physics in the secondary electromagnetic showers, expanding in complementary regions of model parameter space to the higher energy, longer baseline beam dump experiments situated at the NuMI, BNB, or LBNF beams. Future beam dump searches may be possible to fully probe QCD axion parameter space for MeV masses, such as a proposed dump mode or target-less running mode for DUNE~\cite{Brdar:2022vum}. A dedicated target-less mode was shown to test electron-ALP couplings down to $g_{ae}\sim 10^{-6}$ for $m_a < 2 m_e$ and down to $g_{ae}\sim10^{-9}$ from ALP decays to $e^+ e^-$ pairs with a limited 3 month to 1-year exposure. \\

\section*{Acknowledgments}
We are grateful to Ornella Palamara for the helpful discussions regarding the potential for dedicated ALP studies at ArgoNeuT. The work of IMS is supported by DOE under the award number DE-SC0020250. The work of BD and AT is supported by the DOE Grant No. DE-SC0010813. Portions of this research were conducted with the advanced computing resources provided by Texas A\&M High-Performance Research Computing. The work of GG, WJ, and JY is supported by the U.S. Department of Energy under Grant No. DE-SC0011686. We thank the Center for Theoretical Underground Physics and Related Areas (CETUP*) and SURF for facilitating portions of this research.

\bibliography{main}

\appendix

\section{QCD Axion Models}
\label{app:qcd}
The correlations between the QCD axion mass and its effective couplings are given below, taken from ref.~\cite{DiLuzio:2020wdo}. We simply reiterate those correlations here for the convenience of the reader. The relation between the Peccei-Quinn breaking scale $f_a$ and the axion mass is
\begin{equation}
    f_a =  \bigg(\frac{5.691\times 10^{6} \textrm{eV}}{m_a}\bigg) \textrm{GeV}
\end{equation}
To find the correlations between the axion mass and its effective couplings to photons in the Kim-Shifman-Vainshtein-Zakharov (KSVZ) benchmark model~\cite{PhysRevLett.43.103, SHIFMAN1980493} is then given by Eq.~\ref{eq:ksvz_gagamma};
\begin{align}
\label{eq:ksvz_gagamma}
    g_{a\gamma} &= \frac{m_a}{\textrm{GeV}} \bigg(0.203 \frac{E}{N} - 0.39\bigg)
\end{align}
We then consider a range of model parameter space by considering anomaly number ratios of $E/N = 44/3$ to $E/N = 2$. This defines a band in $(m_a, g_{a\gamma})$ parameter space in which the QCD axion's couplings and mass may reside.

For the Dine-Fischler-Srednicki-Zhitnitsky (DFSZ) benchmark model~\cite{Zhitnitsky:1980tq,DINE1981199,Dine:1981rt,Dine:1982ah}, for which couplings to electrons would be dominant relative to the photon couplings, we take
\begin{equation}
\label{eq:dfsz_gae}
    g_{ae} = \dfrac{m_e C_{ae}(m_a, \tan\beta)}{f_a}
\end{equation}
where the coefficient $C_{ae}$ is dependent on the rotation angle $\beta$ for the vacuum expectation values of the extended Higgs sector in DFSZI and DFSZII models;
\begin{align}
    \textrm{DFSZ(I):\, \, }C_{ae} &= -\frac{1}{3} \sin^2\beta  + \textrm{loop factors} \\
    \textrm{DFSZ(II):\, \,  } C_{ae} &= \frac{1}{3} \sin^2\beta + \textrm{loop factors} 
\end{align}
Here we take $\tan\beta$ values between 0.25 and 120, which equates to $\sin\beta = 0.242536$ and  $\sin\beta = 0.999965$, respectively~\cite{Giannotti:2017hny}.

\end{document}